\documentclass[runningheads, 10pt]{llncs}

\usepackage{graphicx}
\usepackage[most]{tcolorbox}
\usepackage[dvipsnames]{xcolor}
\usepackage[font=scriptsize,labelfont={scriptsize,bf}]{subfig}
\usepackage{fontawesome5}
\usepackage{enumitem}
\usepackage{booktabs}
\usepackage{tabularx}
\usepackage{amstext}  
\usepackage{xspace}
\usepackage{amsmath,amssymb}
\usepackage[overload]{empheq}

\usepackage[font=scriptsize]{caption}
\usepackage{graphbox}
\usepackage[colorlinks=true,linkcolor=blue,breaklinks=True,citecolor=brown,urlcolor=blue]{hyperref}
\usepackage{tabularray}
\usepackage{colortbl}
\usepackage{cellspace}
\usepackage{rotating}   
\usepackage{tabularray}
\usepackage{pgfplots}  
\usepackage{dcolumn}   
\definecolor{ACMGreen}{RGB}{0,153,0}
\definecolor{ACMRed}{RGB}{200,0,0}
\newcommand{\cmark}{{\textcolor{ACMGreen}{\ding{51}}\xspace}} 
\newcommand{\xmark}{{\textcolor{ACMRed}{\ding{55}}\xspace}} 

\newcolumntype{?}{!{\vrule width 1.5pt}}

\newcommand{\datasetname}{{\small \texttt{L2DTnH}}\xspace}

\newcommand{\datasetnameunf}{{\texttt{L2DTnH}}\xspace}

\newcommand{\textbox}[1]{
    \noindent\fbox{%
        \parbox{0.97\columnwidth}{%
            {#1}
        }%
    }
}

\newtcolorbox{cooltextbox}[1][]{%
    colback=black!5,
    colframe=black!5,
    notitle,
    sharp corners,
    borderline west={0pt}{0pt}{red!80!black},
    enhanced,
    breakable,
    left=0pt,
    right=0pt,
    top=0pt,
    bottom=0pt
    }

\newcommand\smamath[1]{{\small $#1$}}

\newcommand\revision[1]{%
  \bgroup
  \hskip0pt\color{blue!80!black}%
  #1%
  \egroup
}

\usepackage{amsfonts}
\usepackage{multirow}
\usepackage{diagbox}
\usepackage{setspace}
\usepackage{xurl}
\usepackage{cite}
\usepackage{dcolumn}
\newcommand{\chat}[1]{\detokenize{#1}}
\usepackage{pifont}
\renewcommand{\cmark}{\ding{51}}
\renewcommand{\xmark}{\ding{55}}

\setlist{noitemsep, topsep=0pt,leftmargin=*}

\begin{document}

\title{``bot lane noob'' Towards Deployment of NLP-based Toxicity Detectors in Video Games}

\titlerunning{Towards Deployment of NLP-based Toxicity Detectors in Video Games}

\author{Jonas Ave\inst{1} \and Irdin Pekaric\inst{1} \and Matthias Frohner\inst{2} \and Giovanni Apruzzese\inst{2}\inst{1}}

\institute{University of Liechtenstein, Vaduz, Liechtenstein \and Reykjavik University, Reykjavik, Iceland}

\maketitle 

\begin{abstract}
Toxicity and harassment are widespread in the video-gaming context. Especially in competitive online multiplayer scenarios, gamers oftentimes send harmful messages to other players (teammates or opponents) whose consequences span from mild annoyance to withdrawal and depression. Abundant prior work tackled these problems, e.g., pointing out the negative effects of toxic interactions. However, few works proposed countermeasures specifically developed and tested on textual messages sent during a match---i.e., when the ``harassment'' actually occurs. We posit that such a scarcity stems from the lack of high-quality datasets that can be used to devise ``automated'' detectors based on natural-language processing (NLP) and machine learning (ML), and which can -- ideally -- mitigate the harm of toxic comments during a gaming session.

This work provides a foundation for addressing the problem of toxicity and harassment in video games. First, through a systematic literature review (n=1,039), we provide evidence that only few works proposed ML/NLP-based detectors of toxicity/harassment during live matches. Then, to foster more practical research in this domain, we partner-up with 8 expert League of Legend (LoL) players and create a fine-grained labelled dataset, \datasetname{}, containing 1.4k toxic and 13.8k non-toxic messages exchanged during LoL matches. We use \datasetname{} to develop an ML-based detector that we then empirically show outperforms general-purpose and state-of-the-art toxicity detectors reliant on NLP. Finally, to further demonstrate the practicality of our resources, we test our detector on game-related data beyond that included in \datasetname{}; and we develop a Web-browser extension that flags toxic content in Webpages---without making any query to third-party servers owned by AI companies. We publicly release all of our resources. Our contributions pave the way for more applied research devoted to fighting the spread of toxicity and harassment in video games. \textbf{WARNING: Offensive Language}
\end{abstract}

\section{Introduction}
\label{sec:introduction}
\noindent
Did you know that one out of every three people plays video games~\cite{exploding2025gamers}? Indeed, the attention towards the video-gaming sector is massive: according to 2025 data, the revenue of the video-gaming industry exceeds \$100 billion~\cite{statista2025game}; video games are the most popular type of mobile application~\cite{explodingtopics2025mobile}; and  ''gaming'' accounts for 1/10th of the entire internet traffic (ranked third after ``video'' and ``social''~\cite{explodingopics2025internet}). Video games now play a leading role amidst Web-related technologies.

The ecosystem of modern video games is complex. At a high level, creators of video games---spanning from ``AAA'' studios such as Riot~\cite{riotgames} or Blizzard~\cite{blizzard}, to ``indie'' studios such as SuperGiant games~\cite{supergiant}---make a product for the end users---the players. However, a detailed look reveals that the interactions among these two entities (makers and players) are much more profound. For instance, developers of online multiplayer titles enable players to send messages during a gaming session~\cite{tricomi2023attribute}; gamers also frequently interact via third-party platforms (e.g., Reddit~\cite{redditgaming} or YouTube~\cite{youtube2025games}), some of which entirely (e.g., Steam~\cite{steamcommunity}) or mostly (e.g., Twitch~\cite{twitchstreamers}) depend on game-related content. Unfortunately, such interactions are not always constructive: as abundant reports have pointed out, video games are plagued by the presence of toxicity and harassment~\cite{statista2023harassment,adl2020harassment}.

\vspace{-0.5mm}

A large body of literature has studied the problem of toxicity and harassment in the video game context~\cite{kordyaka2025defining,wijkstra2024tame}. Numerous user studies~\cite{kilmer_addressing_2024,shaer2017understanding,tang2020investigating,beres_dont_2021,zsila2022toxic} highlighted the negative consequences (e.g., withdrawal, self doubt, depression) that such messages can have on targeted players. These works typically advocate for the integration of automated mechanism that can deal with the ``source'' of toxicity/harassment right away---e.g., censorship of the affected messages, or a ban of the offending player. Indeed, recent advances in the machine-learning (ML) and natural-language processing (NLP) domains have the potential to mitigate this problem~\cite{chakrabarty2019machine,d2020bert}. Yet, in the research domain, and to the best of our knowledge, only few works (e.g.,~\cite{martens2015toxicity}) proposed and \textit{implemented} solutions, reliant on ML/NLP, to counter toxicity/harassment in the gaming context. 

\vspace{-0mm}

\textbf{\textsc{Summary of Contributions.}}
We seek to provide a foundation for the \textit{practical development and deployment} of automated detection mechanisms of game-related toxicity and harassment. We make the following contributions:
\begin{itemize}[leftmargin=*, noitemsep]
    \item First, we systematically review prior work (n=1,039) and find that only 15 papers proposed ML/NLP-based mechanisms to counter toxicity/harassment (§\ref{sec:related}). However, such methods always relied on datasets with a limited scope---in terms of openness, source, or labeling granularity. We discuss such limitations.
    \item Then, to overcome such limitations, we create a novel dataset: \datasetname{} (§\ref{sec:dataset}). \datasetname{} draws from the well-known Tribunal dataset~\cite{tribunal}, which reports the chatlogs of League of Legends (LoL) matches that included some (verified) harassment---but without any fine-grained label, making it hardly usable for toxicity detection. So, for \datasetname{}, we recruited 8 expert video gamers and asked them to use their expertise to annotate the ground truth of \textit{each message} in Tribunal. Overall, \datasetname{} has 1,398/13,773 toxic/non-toxic messages---making it the largest open-source and game-specific dataset for toxicity detection.
    \item Next, we show the practicality of \datasetname{} and use it to fine-tune an ML model (§\ref{sec:model}). Comparisons with state-of-the-art transformer models for toxicity detection (e.g., Toxic-BERT) reveal our model outperforms all of them.
    \item Finally, to show the applicability of our tools in the real-world, we tested our model on game data from a different source: captions of YouTube videos (§\ref{sec:plugin}). We also developed a browser extension that operates locally and {\small \textit{(i)}}~does not send any data to remote servers while {\small \textit{(ii)}}~automatically flagging toxic content in Web pages. We discuss lessons learned from our implementation.
\end{itemize}
We release all of our resources~\cite{repository}. We informed Riot Games of our tools.

\section{Related Work and Motivation}
\label{sec:related}
\noindent
We outline the concepts and challenges tackled by our paper~(§\ref{ssec:background}), describe our systematic literature review~(§\ref{ssec:slr}) and finally present the research gap~(§\ref{ssec:focus}).

\subsection{Toxicity and Harassment in Video Games}
\label{ssec:background}
\noindent
``Toxicity'' and ``harassment'' are pervasive in many online communities~\cite{mandryk2023combating,aroyo2019crowdsourcing}. 

To provide some definitions, ``toxicity'' denotes behaviors that are harmful towards other individuals, whereas ``harassment'' indicates targeted and/or repeated instances of toxic behavior, oftentimes resulting in greater harm~\cite{laato_traumatizing_2024}. For instance, harassment can include severe forms of verbal abuse, or sexism~\cite{kilmer_addressing_2024}. Yet, a precise distinction between these two terms is ultimately subjective. Hence, in the remainder of this work, we will consider these two terms as synonyms. 

The video-game context is particularly prone to toxic behaviors, especially within the communities of competitive online multiplayer games---such as League of Legends~\cite{kou2020toxic} (LoL). This specific types of games, occasionally referred to as ``esports'' and counting almost 1 billion users~\cite{statista2024esports}, presents peculiarities that make identification of toxic behaviors through NLP more challenging than in other domains. For instance, terms such as ``noob'' or ``uninstall'' can be used to mock other players, but are meaningless outside a gaming context. At the same time, specific games have their own toxic jargon (e.g., the statement ``Leona is botting'' only has toxic implications within LoL). 
Such specificity makes general-purpose solutions against toxic behaviors (e.g., censoring of bad words) not very effective in the gaming ecosystem, thereby calling for of ad-hoc mitigations.

\subsection{Systematic Literature Review}
\label{ssec:slr}
\noindent
Given the challenges of fighting toxicity and harassment in the video-gaming context, we wondered: ``{\color{purple!50!blue} what prior works proposed automated techniques, based on ML/NLP methods, to detect toxicity/harassment in the gaming context?}'' We tackle such a research question (RQ) via a systematic literature review (SLR).

\textbf{Paper Collection.} Our SLR follows established PRISMA guidelines~\cite{page2021prisma}. The entire procedure was conducted by two authors who interacted and validated each-other's findings. First, between Dec. 2024 and March 2025, we queried four popular databases of peer-reviewed literature (ACM DigitalLibrary, SpringerLink, IEEE Xplore, Elsevier ScienceDirect) for papers matching the queries (``game/player/esport'' \smamath{\land} ``ML/AI/NLP'' \smamath{\land} ``toxicity/harassment'') published since 2014; we perform our searches between Dec. 2024 and March 2025. Such a search yielded a 1,039 papers. The exact search terms are in our repository~\cite{repository}.

\textbf{Screening.} We then manually checked the metadata (title and abstracts) of these papers, removing those that were clearly outside of our scope (e.g., a user study with no solution~\cite{poeller2023suspecting}). Such a process led to excluding 989 papers. The remaining 50 papers were then analysed in their entirety. Such an analysis, led to the removal of 35 papers (e.g., no application of AI~\cite{sparrow2024towards}, or applications with no relevance to NLP~\cite{canossa2021honor}). Hence, out of 1,039 papers, only 15 (\smamath{<}2\%) proposed/implemented some automated mechanisms to address toxicity/harassment.

\textbf{Findings.} Altogether, these 15 papers have provable limitations. Some focus on ``emotes''~\cite{kim2022understanding} or on detecting toxic content not during live gameplay (e.g., Twitch chats~\cite{merayo2024applying,dreier2023toxicity}, YouTube~\cite{obadimu2019identifying}, reviews~\cite{viggiato2021causes}, online forums~\cite{vo2021automatically}). Some do not provide details or do not share the dataset used to test a given hypothesis~\cite{frommel2020recognizing,sengun2019analyzing,stepanova2021natural}. Others~\cite{shannaq2022offensive,cornel2019cyberbullying} do not focus on the English language (i.e., the de-facto language of multiplayer online games~\cite{martens2015toxicity}). 
The most relevant works are:~\cite{blackburn2014stfu,martens2015toxicity,neto2017studying,murnion2018machine}. However,~\cite{martens2015toxicity} only focus on verbal abuse, overlooking other forms of toxic behavior such as trolling; whereas~\cite{murnion2018machine} combine various data sources, including chat logs processed by ML, but focus on a game (World of Tanks) that has long since stopped being popular (e.g., 4k current players in October 2025~\cite{activeplayer2025wot}). Finally,~\cite{blackburn2014stfu} and~\cite{neto2017studying} focus on LoL (still extremely popular~\cite{activeplayer2025lol}) and use the Tribunal dataset which provides aggregated data of entire matches.

\vspace{-0mm}
\subsection{Research Gap and Problem Statement}
\label{ssec:focus}
\vspace{-0mm}
\noindent
Our SLR highlights that
{\small \textit{(i)}}~despite abundant research interest in game-specific toxicity/harassment, {\small \textit{(ii)}}~only few works specifically proposed ML/NLP methods to address this problem within the game itself, and {\small \textit{(iii)}}~previously-proposed contributions have some limitations from a practical viewpoint---as also evidenced by the fact that toxicity/harassment are still an open problem~\cite{statista2023harassment}. 

We argue, echoing the opening statement of~\cite{murnion2018machine}, that one of the reasons why ML/NLP methods have not seen more applications in this context is due to lack of high-quality data. Such a lack prevents both {\small \textit{(a)}}~development of ML/NLP detectors, and {\small \textit{(b)}}~assessment of any sort of mitigation (not necessarily reliant on ML/NLP). By ``high-quality data'' we specifically refer to a large-scale dataset provided with granular ground truth that reflects the specific types of toxic behaviors and which integrates game-specific knowledge. 

Therefore, our primary objective is contribute with a dataset that fulfills the aforementioned requirements. We acknowledge that there are ways to use ML that do not encompass NLP techniques (e.g., analysis of gameplay elements~\cite{canossa2021honor} or audio features~\cite{reid2022bad}). These approaches are valid, but orthogonal to our goal.

{\setstretch{0.99}
\textbox{\textbf{Positive Light.} We do not seek to invalidate prior research. Moreover, due to lack of public code, we cannot reproduce prior contributions. We are merely elucidating factual shortcomings that motivated us to pursue our research objective. Our open-source~\cite{repository} contributions are rooted on prior work.}}

\section{Our Proposed \datasetnameunf{} Dataset}
\label{sec:dataset}
\begin{figure}[t]
  \centering
  \includegraphics[width=0.9\columnwidth]{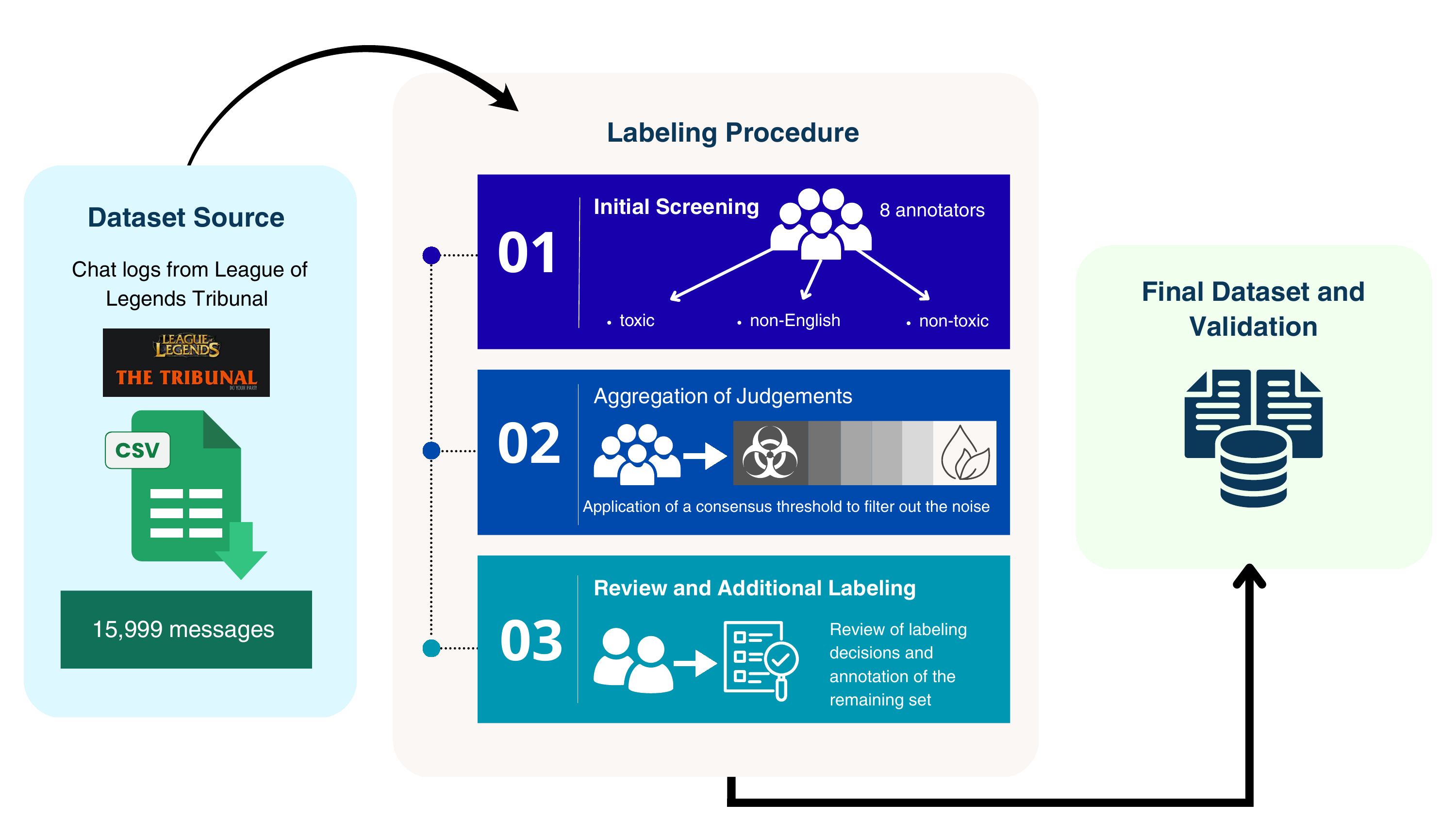}
  \vspace{-3mm}
  \caption{Overview of the creation process of \datasetnameunf{}.}
  \label{fig:dataset}
  \vspace{-5mm}
\end{figure}

\noindent
We describe our proposed \datasetname{}, short for ``\texttt{L}oL-based \texttt{L}abeled \texttt{D}ataset of \texttt{T}oxicity a\texttt{n}d \texttt{H}arassment.'' We first outline the origin and characteristics of the source data (§\ref{ssec:dataset_context}), then present the labeling procedure (§\ref{ssec:re-annotation}), and finally provide quantitative metrics (§\ref{ssec:outcome}). An overview of our methodology is shown in Fig.~\ref{fig:dataset}.

\subsection{Context and Challenges}
\label{ssec:dataset_context}

\noindent 
Our research builds upon the LoL ``Tribunal'' chatlogs dataset~\cite{tribunal}. 
The dataset was created by RIOT Games’ ``Tribunal'' moderation system~\cite{riottribunal}, an initiative where experienced players collectively review matches wherein an player had been reported to exhibit toxic behavior---with the purpose of verifying if such claims are true. Each case in the Tribunal dataset contains complete in-game chatlogs and metadata (e.g., player roles, champions, team), including the decision obtained by majority voting among reviewers. 

The Tribunal dataset is valuable because it contains large-scale ($>$1M chatlogs) real-world communications between players in a highly competitive online environment that have been verified as being toxic. Unlike social media platforms or forums, LoL chat contains fast-paced exchanges, fragmented utterances, as well as strong emotional reactions. The dataset thus contains player interactions that include sarcasm, slang, and game-specific abbreviations that are rarely found in general-purpose toxicity datasets (e.g.,~\cite{wijesiriwardene2020alone}).

Despite such advantages, the Tribunal dataset has a notable shortcoming: \textbf{annotations are provided only at the match level}.
Indeed, if a player in a match is deemed ``toxic'', then the entire match is considered as ``toxic''. 
This means that all messages from a toxic match are considered toxic---even neutral or supportive ones (e.g., messages such as ``w8'' or ``ok'' are treated as ``offensive language'' simply because they appear in a toxic match). In other words, the information provided in the Tribunal dataset does not allow to identify the specific toxic messages. Such a limitation hence inhibits carrying out message-level classification of toxic content.

\begin{table}[t]
\centering
\scriptsize 
\setlength{\tabcolsep}{3pt}
\caption{Annotator background and labeling contribution.}
\label{tab:annotators}
\begin{tabular}{lccccc}
\toprule
ID & Gaming Exp. & Rank & Hrs/wk & Other Exp. & Msgs \\
\midrule
A1 & 17 yrs  & Masters    & 17 & MOBA, FPS, MMOS      & 5k \\
A2 & 20 yrs & Diamond & 21 & MOBA, FPS, MMOS            & 12k \\
A3 & 9 yrs  & Diamond & 30 & MOBA, SBX, FPS         & 5k \\
A4 & 8 yrs  & Gold    & 28 & MOBA, SBX, FPS          & 5k \\
A5 & 6 yrs  & Diamond  & 19  & MOBA, SBX, FPS         & 5k \\
A6 & 7 yrs  & Bronze  & 10 & MOBA, SBX, FPS       & 15k \\
A7 & 20 yrs  & Diamond  & 20 & MOBA, SBX, FPS       & 5k \\
A8 & 9 yrs  & Silver  & 35 & MOBA, MMORPG       & 15k \\
\bottomrule
\end{tabular}
\vspace{-6mm}
\end{table}

\subsection{Fine-grained Annotation}
\label{ssec:re-annotation}
\noindent
We carried out our annotation to converts the original ``match-level'' Tribunal dataset into a ``message-level'' resource, i.e., our proposed \datasetname{} dataset.

\textbf{Annotator selection}
\label{sssec:annotator}
We recruited eight annotators who were all long-term LoL players with 6--20 years of experience in the gaming domain. Table~\ref{tab:annotators} summarizes their backgrounds, including average LoL weekly playtime and in-game rank. Prior familiarity with the specific game’s linguistic culture was a key selection criterion, as gaming toxicity is frequently expressed through linguistic cues such as sarcasm (``nice ult bro''), abbreviated insults (``ez''), or creative misspellings (``n0000b''). Annotators unfamiliar with these conventions can misinterpret intent, introducing substantial bias. 

\textbf{Labeling procedure}
Labeling each message of the over 1M chatlogs within the Tribunal dataset is not humanly possible, especially because determining whether any given message is toxic or not is ultimately subjective~\cite{lobo2022supporting,braun2024understanding}.\footnote{Similar challenges in creating labeled datasets are previously reported (e.g., \cite{schroer2025dark}).} So we devised a best-effort strategy founded on consistency, organized in three steps.
\begin{itemize}[leftmargin=*]
    \item[i)] First, each annotator was assigned the same set of 5,000 messages,\footnote{We took the first 5k messages in Tribunal, which can be considered as random sampling since the distribution of toxicity across Tribunal can be assumed as random.} and was tasked to independently assign the label ``toxic,''\footnote{E.g., insults, harassment, offensive tone, or otherwise suspicious content.} ``non-toxic,''\footnote{In principle, ambiguous cases such as sarcasm or playful teasing are toxic only if a player would reasonably perceive them as offensive within a competitive match.} or ``non-English''. Annotators were explicitly instructed to base judgments solely on linguistic cues---potentially accounting for quick repeated messages sent by the same player in a short timeframe (e.g., ``yo'', ``are'', ``trash''). 
    \item[ii)] Then, to handle subjectivity, we aggregated the judgments by applying a consensus threshold: a message was deemed as ``toxic'' if at least two annotators labeled it as such (see Table~\ref{tab:dataset_example_labeled_reviewers}).\footnote{This threshold was determined after continuous assessments, and we occasionally even inquired the specific annotator(s). Ultimately, toxicity is subjective and if two people agree a message has toxic potential, then it should be deemed as such.} Such a procedure enabled to filter out noise, while ensuring that a ``toxic'' message could be truly perceived as toxic.
    \item[iii)] Finally, three annotators reviewed the labeling decisions and proceeded to annotate an additional set of 10,999 messages (one only participated in 7k). For this step, disagreements or borderline cases were discussed jointly until a final consensus was reached.
\end{itemize}
Our annotation hence increased the number of labeled instances by 240\% without demanding additional human-working hours by five of the eight annotators beyond that of the first 5k messages---which have been used to ``align'' the mindset of the last three annotators with that of entire set of annotators.

\textbf{Validation metrics} 
To objectively validate our annotation strategy, we computed well-known statistics after the second step.
The Fleiss' \smamath{\kappa=0.62}, indicates substantial agreement among the eight annotators~\cite{falotico2015fleiss}.
Overall, 4,704 messages (out of 5,000) were unanimously marked as ``toxic'', ``non-toxic'', or ``non-English'' demonstrating strong consensus on highly-abusive expressions. Hence, the labeling was consistent across our sample.
Finally, \smamath{\approx} 1.9\% messages were flagged as toxic only by a single annotator, and hence resolved as non-toxic.

\begin{table}[!t]
\centering
\scriptsize
\caption{Examples of annotated chat messages with decisions.}
\label{tab:dataset_example_labeled_reviewers}
\setlength\tabcolsep{2pt} 
\begin{tabular}{ccccccccccc}
\toprule
\textbf{ID} & \textbf{Message} & A1 & A2 & A3 & A4 & A5 & A6 & A7 & A8 & \textbf{Label} \\
\midrule
37  & \chat{mother fucking noob}                       & 1 & 1 & 1 & 1 & 1 & 1 & 1 & 1 & 1  \\ 
83  & \chat{rammus is such a slut}                      & 1 & 0 & 0 & 0 & 0 & 1 & 1 & 1 & 1 \\
87  & \chat{FUCK}                                       & 1 & 0 & 0 & 0 & 0 & 0 & 0 & 0 & 0 \\
114 & \chat{get fked}                                   & 1 & 1 & 0 & 0 & 1 & 1 & 1 & 1 & 1 \\
123 & \chat{rammus useless}                             & 0 & 1 & 0 & 1 & 0 & 1 & 0 & 1 & 1 \\
172 & \chat{leona you are so bad. you lost this lane gg} & 1 & 1 & 0 & 0 & 0 & 1 & 0 & 1 & 1 \\
203 & \chat{bot lane noob}                              & 1 & 1 & 0 & 0 & 0 & 1 & 0 & 1 & 1 \\
213 & \chat{fucking scumbags blocked the lantern}       & 1 & 1 & 0 & 0 & 1 & 1 & 1 & 1 & 1 \\
257 & \chat{na u suck lucain}                           & 1 & 1 & 0 & 0 & 1 & 1 & 1 & 1 & 1 \\
315 & \chat{just uninstall lol}                         & 1 & 1 & 0 & 0 & 0 & 1 & 1 & 1 & 1 \\
\bottomrule
\end{tabular}
\vspace{-6mm}

\end{table}

\subsection{Quantitative Analysis}
\label{ssec:outcome}

\noindent
At the end of our annotation procedures, we obtained a set of 15,999 labeled messages, representing the chatlogs of \smamath{\approx}100 different LoL matches containing some (verified) instances of toxicity.

Specifically, out of 15,999 messages, 1,398 (8.74\%) are ``toxic'', 13,773 (86.09\%) are ``non-toxic'' and 828 (5.17\%) are ``non-English''. Across the 15,171 English messages, the average length (in characters) is 12.5 (std=11.53).

We can hence approximate the overall ``cost'' of our labeling efforts. A conservative assumption is that each message required 5 seconds to be labeled (which includes: reading the message, and potential surrounding messages; determining its ground truth; and physically assigning the label). Hence, the first step of our labeling procedure required: 5 seconds \smamath{\times} 8 annotators \smamath{\times} 5,000 messages = 200k seconds \smamath{\sim} 56 hours; whereas the third step required: 5 \smamath{\times} (2 \smamath{\times} 10,999 + 7,000) = 145k seconds \smamath{\sim} 40 hours. Taking into account discussion and reviewing, we can hence estimate that \datasetname{} required \smamath{\approx}100 human-working hours to be created.

\section{Using \datasetnameunf{} in Practice: Empirical Tests}
\label{sec:model}
\noindent
It is factual that our \datasetname{} represents a valid testing ground for toxicity detectors. However, a question arises: ``{\color{purple!50!blue} Does \datasetname{} improve existing ML- and NLP-based toxicity detectors?}'' We answer this question with an original assessment of various models reliant on pre-trained transformers. We first describe the experimental setup~(§\ref{ssec:testbed}), then present the results (§\ref{ssec:results}); finally, we examine the effects of using \datasetname{} at different levels of message grouping (§\ref{ssec:game_level_analysis}).

\subsection{Testbed and Model Development}
\label{ssec:testbed}

\noindent
An investigation of the current landscape of pre-trained transformers reveals that Toxic-BERT~\cite{Detoxify} is among the best general-purpose models for toxicity detection, so we consider this as a baseline. 

To tailor this model to the gaming domain, we fine-tune it on our \datasetname{}. Thus, we take the 15,171 English messages in \datasetname{} and use stratified sampling to partition them into a training and testing set by applying an 80:20 split. Fine-tuning was implemented using the Hugging Face Transformers and PyTorch frameworks. The model’s original six-class classification head was replaced by a binary classification layer. Input sequences were tokenized with the BERT tokenizer (maximum sequence length = 192 tokens). Training was performed for four epochs with a batch size of 64, using the AdamW optimizer (learning rate = $2e-5$) and a linear learning-rate scheduler. Early stopping with a patience of two epochs was applied to prevent overfitting. We set the CrossEntropyLoss as the objective function. Implementation details are in our repository~\cite{repository}.

At the end of this fine-tuning procedure, we obtain a new model which we denote IGC-BERT (Inappropriate Game Chat-BERT).  

\subsection{Evaluation Results and Comparison}
\label{ssec:results}
\noindent
For a broad a comprehensive assessment, we compare our IGC-BERT against state-of-the-art and publicly-accessible models.

Specifically, we consider: \textit{protectai\_unbiased\_toxic\_roberta}~\cite{protectai}, \textit{nicholas\_kluge\_\allowbreak toxicity\_model}~\cite{nicholas22aira}, \textit{junglelee\_bert\_toxic\_comment\_} \textit{classification}~\cite{junglelee}, \textit{martin\_ha\_\allowbreak toxic\_comment\_model}~\cite{martinha}, \textit{garak\_llm\_} \textit{roberta\_toxicity\_classifier}~\cite{logacheva-etal-2022-paradetox}, and the baseline \textit{unitary\_toxic\_bert} (as an ablation study). All such models are adapted to a binary classification setting. The assessment is always done on the same test set (i.e., 20\% of \datasetname{}). The entire codebase is in our repository~\cite{repository}.

We also considered two large-language models: first, \textbf{ChatGPT} and, specifically GPT-4o (the June 2025 version). We used its API to make our tests, which required us to purchase API queries and forced us to carry out a smaller evaluation on a subset of 667 non-toxic and 40 toxic messages (\smamath{\approx}1/3 of our test set). Given its multi-purpose focus, we devised two one-shot prompts for assessing ChatGPT: 
P1=``{\small You are a classifier that detects toxic behavior in gaming chats. Classify the following message as either `toxic' or `non-toxic'. The message may contain slang, sarcasm, abbreviations, or profanity. Your response must be only: toxic OR nontoxic.}'' and 
P2=``{\small You are a classifier that detects if a message in gaming chats is inappropriate. Classify the following message as either `inappropriate' or `non-inappropriate'. The message may contain slang, sarcasm, abbreviations, or profanity. Your response must be only: inappropriate OR non-inappropriate.}'' Hence, we made \smamath{\approx}1,400 queries to OpenAI's API, and recorded the answers from GPT-4o. Then, we considered another LLM: the (free) \textbf{Llama~3.2}, using a variant of P1 (reported in our repo~\cite{repository}) which we adjusted for the characteristics of Llama~3.2. 

The results of our assessment are shown in Table~\ref{tab:model_comparison}, reporting Accuracy, Precision, Recall, and F1-score (we consider a ``positive'' as a toxic message). For completeness, the test set has 2,755 non-toxic and 280 toxic messages. We appreciate that fine-tuning on our \datasetname{} yielded statistically significant improvements (validated with a t-test at \smamath{p>.05}) over the baseline model, with improvements of nearly 20 absolute percentage points in the F1-score, and the false positives decreased from 137 to 32. We investigated these results: the baseline model classified neutral or sarcastic expressions as toxic, due to lack of adaptation to the game/LoL-specific domain. ChatGPT 4o and LLama 3.2 do not seem to be very effective, as indicated by underwhelming precision (always below 0.4).

\begin{cooltextbox}
    \textbf{\textsc{Takeaway.}} Fine-tuning on our proposed \datasetname{} leads to statistically-significant (\smamath{p<.05}) improvements: our IGC-BERT model outperforms existing general-purpose toxicity detectors (thanks to our labeled dataset).
\end{cooltextbox}

\begin{table}[t]
\caption{Comparative performance of state-of-the-art toxicity detection models on the test portion (20\%) of our \datasetnameunf{} dataset. For cost reasons ChatGPT was tested on a subset of our test set.}
\label{tab:model_comparison}
\centering
\renewcommand{\arraystretch}{0.9} 
\setlength{\tabcolsep}{1pt}       
\scriptsize                           
\begin{tabular}{lcccc}
\toprule
\textbf{Model} & \textbf{Acc.} & \textbf{Prec.} & \textbf{Rec.} & \textbf{F1} \\
\midrule
protectai\_unbiased\_toxic\_roberta\_onnx        & 0.9021 & 0.4659 & 0.4143 & 0.4386 \\
nicholas\_kluge\_toxicity\_model                 & 0.8708 & 0.3866 & 0.6821 & 0.4935 \\
junglelee\_bert\_toxic\_comment\_classification  & 0.8722 & 0.3977 & 0.7500 & 0.5198 \\
martin\_ha\_toxic\_comment\_model                & 0.9068 & 0.4908 & 0.2857 & 0.3612 \\
garak\_llm\_roberta\_toxicity\_classifier        & 0.9074 & 0.4978 & 0.3964 & 0.4414 \\
ChatGPT 4o (June 2025) P1 & 0.9008 & 0.3495 & 0.9231 & 0.5070 \\
ChatGPT 4o (June 2025) P2 & 0.9178 & 0.3662 & 0.6667 & 0.4727 \\
Llama 3.2 & 0.7542 & 0.2451 & 0.8000 & 0.3752 \\
\textbf{unitary\_toxic-bert (base)}               & 0.9166 & 0.5401 & 0.6500 & 0.5900 \\
\textbf{IGC-BERT (fine-tuned)}                   & \textbf{0.9605} & \textbf{8571} & \textbf{0.6857} & \textbf{0.7619} \\
\bottomrule
\end{tabular}
\vspace{-6mm}
\end{table}

\subsection{Examining different aggregation techniques}
\label{ssec:game_level_analysis}

\noindent 
We study the effects of using our proposed \datasetname{} by accounting for different ways to aggregate the messages contained therein.

\textbf{Rationale.}
Toxic communication in multiplayer games rarely occurs in isolation. Short, fragmented chat messages accumulate into larger sequences that reflect emotional escalation or interpersonal conflict. To capture these contextual effects, we scrutinize our fine-tuned IGC-BERT model at three hierarchical levels of granularity: {\small \textit{(i)}}~\textit{message level}---single chat entries evaluated independently; {\small \textit{(ii)}}~\textit{grouped-message level}---consecutive messages from the same player in a short time frame combined into a single utterance; {\small \textit{(iii)}} \textit{match level}---aggregation of all chat messages produced by one player during an entire game.

\textbf{Setup.} For a fair and consistent evaluation, we followed a slightly different workflow than that presented in §\ref{ssec:testbed}, which assumed a random split. This is because we cannot create chains of ``grouped messages'' (from the test set) if each message is drawn randomly. So, for this assessment, we create the training set by considering all messages exchanged in 81 games (i.e.,\smamath{\approx}80\% of all matches in \datasetname{}) and the test set by considering all matches of the remaining 18 games (i.e., \smamath{\approx}20\% of all matches in \datasetname{}). We hence used the training set to create another fine-tuned variant of our IGC-BERT.

\textbf{Results.} We report the results  in Table~\ref{tab:granularity_comparison}. Specifically, for the message-level results, we simply tested on the (new) test set (in the same fashion as in §\ref{ssec:results}) and we see that the performance aligns with that shown in the last row of Table~\ref{tab:model_comparison}: small differences in the recall can be explained with a different distribution of toxic messages in these games, which are not captured during the fine-tuning of this variant of IGC-BERT. For the Grouped-message level results, we fairly aggregate messages of the same player within a short timeframe (e.g., 10s) and input such ``longer'' messages to the model: we see a substantial increase in the performance (i.e., the recall increases by 12 absolute percentage points) because the model can better recognize toxic content which would be otherwise missed by few-word messages. For the Match-level results, we aggregate all messages of the same player together and submit it to the model. Such an evaluation yields a very high precision with almost no false-positives (only 2 players have been falsely flagged as being toxic), because the model is always able to pinpoint at least some instances of toxicity within the (very long) string received as input. 

\begin{cooltextbox}
    \textbf{\textsc{Takeaway.}} By fine-tuning BERT on the individually-labeled messages in \datasetname{}, one can develop a model that precisely detects toxic players in a match.
\end{cooltextbox}

\begin{table}[t]
\caption{Performance of IGC-BERT across different contextual granularities of message aggregation (note: the IGC-BERT model here follows a different fine-tuning process than that in Table~\ref{tab:model_comparison})}
\label{tab:granularity_comparison}
\centering
\renewcommand{\arraystretch}{0.9} 
\setlength{\tabcolsep}{4pt}       
\scriptsize                           
\begin{tabular}{lcccc}
\toprule
\textbf{Evaluation Level} & \textbf{Acc.} & \textbf{Prec.} & \textbf{Rec.} & \textbf{F1} \\
\midrule
Message level          & 0.9605 & 0.8571 & 0.6857 & 0.7619 \\
Grouped-message level   & 0.9712 & 0.8974 & 0.8065 & 0.8491 \\
Match level             & \text{0.9157} & \text{0.9701} & \text{0.8442} & \text{0.9028} \\
\bottomrule
\end{tabular}
\vspace{-6mm}
\end{table}

\section{Beyond\,the\,dataset: using our tools on\,the Web}
\label{sec:plugin}
\noindent
Insofar, we have evaluated existing (and new) models on our proposed \datasetname{} dataset. Here, we further demonstrate the practical value of our contributions by exploring ``unknown'' game-related contexts. First, we test our IGC-BERT model on LoL videos on YouTube~(§\ref{ssec:youtube}). Then, we integrate IGC-BERT in a custom-made browser extension~(§\ref{ssec:extension}). Finally, we compare \datasetname{} against other datasets for toxicity detection~(§\ref{ssec:other}).

\subsection{Testing IGC-BERT on LoL YouTube Videos}
\label{ssec:youtube}
\noindent
We find it instructive to evaluate the performance of IGC-BERT on YouTube videos entailing LoL-related content.

\textbox{\textbf{Disclaimer.} To avoid fingerpointing exercises, we deliberately chose videos that are sarcastic and/or whose creators are well-aware that the video includes some form of toxicity. We merely seek to gauge if our model can detect possible toxic instances in captions---which are semantically different than chatlogs.}

\vspace{1mm}

\textbf{Why is it relevant?}
We designed our \datasetname{} to capture instances of toxicity/harassment during a LoL match. However, similar instances can occur also outside of a match. For instance, YouTube is a popular resource for LoL-related content: players can upload videos of their games, and share them on the Web. Unfortunately, some videos can include toxic content, e.g., imprecations of the creator after being defeated by an enemy, or ``trash talking'' a certain opponent with game-specific jargon. Even though such occurrences may not reach the actual player, downstream viewers can be hurt by such content, or can find it disrespectful---which can lead to, e.g., the video being reported or taken down. Therefore, content creators can use our resources to automatically analyse their videos \textit{before} uploading them, thereby preventing harmful consequences.

\textbf{Methodology.} 
Our goal is simple: assessing the extent to which our IGC-BERT model can detect toxic content in LoL videos uploaded on YouTube. We note that our IGC-BERT model expects \textit{textual data} as input. So, to enable our model to analyse YouTube videos, we download the \textit{captions} (automatically generated by YouTube) and submit them, line-by-line, to our IGC-BERT model. However, and unfortunately, we are not aware of any publicly-available dataset of YouTube-videos' captions that contain verified toxic instances of LoL-related gameplay. Hence, our assessment is just a proof-of-concept experiment, meant to demonstrate an ancillary (but practical) application of our resources to address toxicity/harassment in the gaming context.
Therefore, we identify a total of 9 YouTube videos which we expect may contain some toxic content (we found them by submitting the query ``lol toxicity'' on YouTube), retrieved their captions, submitting them to our IGC-BERT model (developed in §\ref{ssec:testbed}) and analysed if our model could classify any line of the captions as ``toxic'', thereby indicating that the audio of the YouTube video has toxic content.

\vspace{1mm}

\textbf{Results.} We report below the 9 videos (clicking on the title leads to the video on YouTube) and an exemplary ``toxic'' captions line (according to IGC-BERT).
{\scriptsize
\begin{enumerate}[leftmargin=*]
    \item \href{https://www.youtube.com/watch?v=jVCBwPxBqtE}{Best Trolls of 2021  League of Legends} ``i hear your trash''

    \item \href{https://www.youtube.com/watch?v=1l4zvEwE6AU}{League of Legends is somehow getting MORE Toxic}  ``this bondage boy sex ring jail escap who''

    \item \href{https://www.youtube.com/watch?v=iCU4otr-Nms}{Toxic ADC flames me... but he doesn't know I'm the Rank 1 Senna} ``completely worthless baby raging ADC but''

    \item  \href{https://www.youtube.com/watch?v=s8bwaEbNNDs}{Why League Is The MOST TOXIC Game of All Time}  ``How can you all be this fed and cry ''

    \item  \href{https://www.youtube.com/watch?v=T-6Jkw2A8iA}{EUW IS THE MOST TOXIC SERVER IN LEAGUE OF LEGENDS HISTORY}  ``vagar you little bell pepper you even''

    \item  \href{https://www.youtube.com/watch?v=PNCSHTD6ywo}{I wanted to quit... \#10}  ``right you are messed up in the head''

    \item  \href{https://www.youtube.com/watch?v=Fn5I-jTI9YM}{The most toxic player in League \#8}  ``cannon no a jungler has mental''

    \item  \href{https://www.youtube.com/watch?v=2VjLXObhtBc}{Why League of Legends is SO TOXIC \_ League of Legends}  ``game because you're all trash stay''

    \item  \href{https://www.youtube.com/watch?v=2VjLXObhtBc}{Why League of Legends' Design Encourages Toxicity \_ Design Delve}  ``some dude saying YOUR MUM, FAT BUM, WIDE TUM''
\end{enumerate}
}
\vspace{-0mm}
Such a proof-of-concept experiment indicates that our IGC-BERT model can be applied to detect toxicity/harassment in LoL-related YouTube videos. Such a mechanism can not only be used by content creators to preemptively issue warnings to their audience, or reconsider uploading their video, but can also be integrated in video-sharing platforms (not limited to YouTube) to inform their users (creators or watchers) that a given video may contain toxic content. 
N.b.: the added value of IGC-BERT is that it flags toxic content pertaining to game-specific jargon---which is seldom captured by general-purpose toxicity detectors.

\subsection{Development of a Browser Extension}
\label{ssec:extension}
\noindent
We developed a browser extension that uses our IGC-BERT model to automatically censor certain elements of Web pages that are flagged as toxic. We first explain why we did so, then outline the design objectives and the implementation choices and presenting the technical requirements of our extension.

\textbf{Motivation}
\label{sssec:extension_motivation}
Using a browser extension to automatically detect toxic content is not new. However, in October 2025, we carried out a literature review by querying Google Scholar with the strings (``browser extension'' \smamath{\land} ``harassment/toxicity''), considering the first 100 results of each query. 
We found only 8 peer-reviewed papers that truly attempted practically implement a browser extension for toxicity detection. Unfortunately, most of these (i.e.,~\cite{bowker2022reducing,bonthu2024civilitycheck,pawale2025toxiguard,sikiandani2025browser,deep2024creating}) do not release their source code, preventing reproducibility. Others (e.g.,~\cite{kwon2018tweety,jahanbakhsh2022our}) do not focus on toxicity/harassment, or not on the English language (e.g.,~\cite{sikiandani2025browser}) or not on the gaming domain (e.g.,~\cite{kazimierczak2023enhancing}). Such a landscape motivated us to develop our extension and share our lessons learned in this paper.

\begin{figure}[t]
    \centering
    \includegraphics[width=0.8\linewidth]{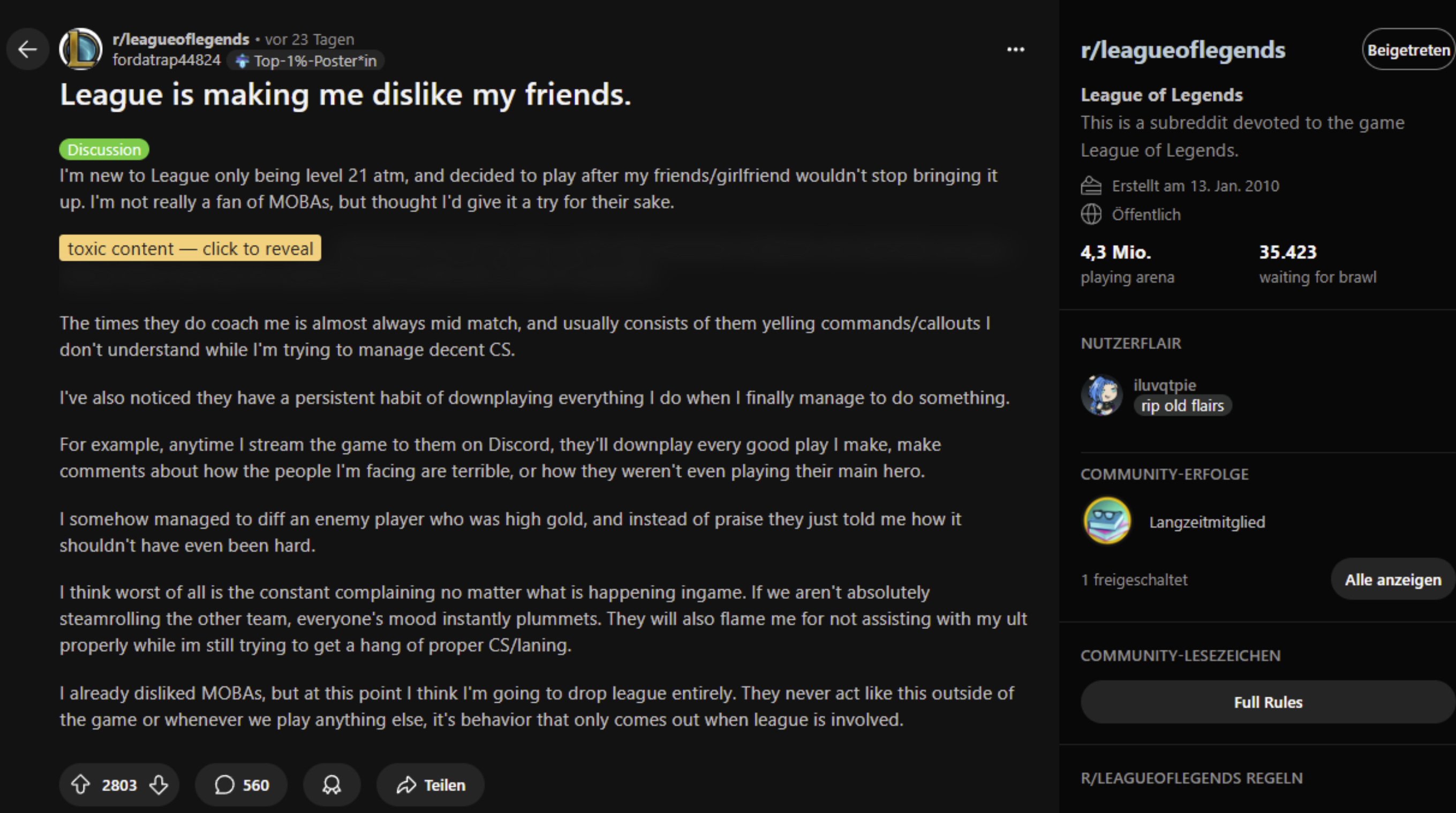}
    \vspace{-3mm}
    \caption{Exemplary application of our Browser Extension (website:~\cite{extension})}
    \label{fig:screenshot}
    \vspace{-7mm}
\end{figure}

\textbf{Objectives}
\label{sssec:extension_objectives}
Our extension has one purpose: analyse web pages, and proactively censor any content that is toxic, preventing the end-user from reading harmful messages. We envisioned such a process as a two-step approach: first, when a user lands on a webpage, the HTML is sent to the extension; then, the extension analyses the HTML via our IGC-BERT model and, if it detects toxic content, it conceals it under a ``spoiler'' tag that can be removed at user discretion (see Fig.~\ref{fig:screenshot}).
The extension works entirely in-browser: we explicitly forbid executing any sort of query to remote servers with the purpose of, e.g., having powerful ML-based models analyse the Web pages browsed by the user and flag potential toxic elements (as done, e.g., in~\cite{bowker2022reducing}, which uses Perplexity AI)).

\textbf{Implementation}
\label{sssec:extension_implementation}
We explain the technical implementation of our extension.
\begin{itemize}[leftmargin=*]
    \item All inference is performed locally in the browser (no cloud connections, usable offline). For implementation, \texttt{web\_accessible\_} \texttt{resources} are used for \texttt{vendor/*}, \texttt{models/}, \texttt{ml/}. Dynamic import is performed using {\small\texttt{import(chrome.\allowbreak runtime.getURL(...))}} in the content script and a local interface ({\small \texttt{initIfNeeded()}, \texttt{runDetectorBatch()}}) in {\small \texttt{ml/inference.js}}. As a result, content does not leave the page, and the extension works without an internet connection. 

    \item A quantised ONNX model~\cite{ducasse2021benchmarking} was used for the model implementation. This choice was made due to the reduced memory requirements and shorter loading times, as well as faster inference compared to non-quantised models. 
    Indeed, impossibility to rely on external API calls meant that all computing had to be done locally, so it was crucial to minimise resource expenditure. 

    \item The inference of content in a web page was batch processed to reduce hang-ups. 
    The implementation was carried out with {\small \texttt{runDetectorBatch()}} using the configuration parameters of our custom IGC-BERT model. The result was a more stable user interface, fewer system failures and consistent progress.

    \item An issue we encountered was handling of abortions, which triggered duplicate scans. Such occurrences were frequent when changing tabs quickly changing webpages. So, we used a flag ({\small \texttt{aborted=true}}) that would trigger an early stopping of the scanning of the webpage if necessary.
\end{itemize}
Since the scan occurs when the page loads, the extension is not designed to work on content that loads progressively (e.g., the news-feed of popular online social networks). Dynamic scanning substantially increases the computational load.

\textbf{Technical requirements}
\label{sssec:technical}
Our final version of the extension (for which we provide its interface in Fig.~\ref{fig:interface}) requires 545MB of free space, most of which (95\%) are for storing the model. We tested the extension on a ``heavy'' webpage with a lot of text (i.e.,~\cite{extension}). Earlier version of the extensions required 4GB of RAM and took more than 5m to process everything and resulted in crashing the browser; the most recent version takes less than 2m to do so (with no crashes). We monitored the memory utilization during the processing of the testing webpage: in ``idle'' (i.e., with the default splash screen), the browser absorbs 520MB, which raises to 780MB after landing on~\cite{extension}; during the 2m of the scan, the used memory increases to 1,400MB; after the scan ends, the used RAM drops to 1,300MB. These tests have been done on a machine equipped with an AMD Ryzen 9 7950 (the quantised IGC-BERT model runs on CPU) and 32GB of RAM.

\vspace{-1mm}
\begin{cooltextbox}
    \textbf{\textsc{Takeaway.}} Development of a browser extension that integrates our model is possible, but processing heavy webpages may require long loading times. By offloading the computation to third-party servers, such analysis would be faster---but such responsiveness would sacrifice privacy.
    We provide in our repo~\cite{repository} a \href{https://github.com/irdin-pekaric/esorics26_toxicity/demo.mp4}{30s demo} of our extension. Moreover, we show in Fig.~\ref{fig:screenshot} a screenshot of our extension, which blocked some toxic content in a LoL subreddit.
\end{cooltextbox}

\subsection{Other Datasets for Toxicity Detection}
\label{ssec:other}
\noindent
Our \datasetname{} is not the only dataset usable for game-related toxicity detection. Let us fairly compare our proposed dataset with those used by prior work.

To this end, we consider the datasets we found mentioned in the papers we analysed during our SLR (in §\ref{ssec:slr}). For each of these datasets, we consider: whether they are publicly accessible; their number of samples; the data source (e.g., forum posts, or chat messages); the granularity; the game/platform from which they are taken. We report the results of our analysis in Table~\ref{tab:datasets}.

Most prior works considered datasets that are not publicly available. Three works (i.e.,~\cite{blackburn2014stfu,kwak2015exploring,neto2017studying}) considered \textit{subsets} of Tribunal which are not released, and the artifacts are not accessible (as of March 2026) preventing reproducibility. Among those for which the dataset is publicly available, we mention: ``Corpus-Twitch-Videogames'' and the custom one in~\cite{dreier2023toxicity}, containing chat messages sent on Twitch (i.e., not during a match); and the Cyberbullying dataset, which refers to LoL and WoW, but contains forum messages. Hence, to the best of our knowledge, our proposed \datasetname{} is the largest publicly-available dataset containing fine-grained and entirely manually-labeled messages\footnote{E.g., the recent GameTox dataset~\cite{naseem2025gametox}, which is unrelated to LoL and was not captured in our SLR, is larger but labeling relied mostly on LLMs which are error-prone.} sent during LoL in-game matches, and which are valid for toxicity detection.

We also mention the existence of other datasets for toxicity detection (e.g.,~\cite{hartvigsen2022toxigen,ghosh2022detoxy,costa2024mutox}), but not meant for the video-game context.

\section{Discussion}
\label{sec:discussion}
\noindent
We outline the limitations of our research (§\ref{ssec:limitations}), discuss lessons learned (§\ref{sssec:implications}), and make some ethical considerations~(§\ref{sssec:ethics}).

\subsection{Limitations, Disclaimers, and Threats to Validity}
\label{ssec:limitations}
\noindent
Our overarching goal is to spearhead development of practical solutions for countering toxicity/harassment in the gaming context---which are needed today. 

We acknowledge that our SLR presents the limitations of querying databases. For instance, our search results did not include a related work~\cite{yang2024game}, which surprisingly was not returned by the ACM DL (despite matching our search terms). Still, the paper~\cite{yang2024game} is complementary to ours, as it focuses on different games (For Honor, and Rainbow Six Siege) and leverages proprietary data. 

The creation of our \datasetname{} (in §\ref{sec:dataset}) was done by finding a trade-off between labeling quality and dataset size. The original Tribunal dataset has millions of messages and labeling all of them is beyond the capabilities of a single research unit. Still, our results (in §\ref{sec:model} and in §\ref{sec:plugin}) show improvement over the baseline and utility both within our dataset and in unknown (but related) domains. Importantly: we do not claim that \datasetname{} can be used for toxicity detectors of \textit{any} game: \datasetname{} contains data pertaining to LoL, so its applicability beyond LoL is uncertain. Therefore, we do not find any threat to the validity of our conclusions concerning the utility of our proposed \datasetname{} for its intended purpose.

The experiments done in unknown domains are a proof of concept, and our browser extension is just a prototype (§\ref{sec:plugin}). We do not claim that our resources can be deployed in operational systems. Also, we underscore that the superior performance of IGC-BERT over general-purpose models (see Table~\ref{tab:model_comparison}) is due to fine-tuning on our proposed \datasetname{}, and not due to methodological novelty.

\begin{table}[t]
  \centering
  \caption{Datasets Used in Game-Related Toxicity Research}
  \label{tab:datasets}
  \scriptsize
  \renewcommand{\arraystretch}{0.9}
  \begin{tabular}{ccccccc}
    \toprule
    \textbf{Dataset} & \textbf{Access} & \textbf{Gran.} & \textbf{Size} & \textbf{Source} & \textbf{Game/Platform} & \textbf{Used in} \\
    \midrule
    Corpus-Twitch & \href{https://github.com/noemeralv/Corpus-Twitch-Videogames}{\cmark{}} & message & 2.6k & chat msgs. & Twitch & \cite{merayo2024applying} \\
    WotReplays & \xmark & message & 15k & replays & WOT & \cite{murnion2018machine} \\
    Custom Dataset & \xmark & phrase & 765k & chat msgs. & RO, DOTA & \cite{cornel2019cyberbullying} \\
    Custom Dataset & \href{https://github.com/noemeralv/Corpus-Twitch-Videogames}{\cmark{}} & message & 100k & chat msgs. & Twitch & \cite{dreier2023toxicity} \\
    Dotalicious & \xmark & word & 7M & chat msgs. & DOTA & \cite{martens2015toxicity} \\
    Cyberbullying dataset & \href{https://ub-web.de/research/}{\cmark{}} & comment & 34k & forum & LoL, WoW & \cite{vo2021automatically} \\
    Tribunal dataset & \xmark & report & 11M & chat msgs. & LoL & \cite{kwak2015exploring} \\
    RIOT dataset & \xmark & thread & 89 & forum & LoL & \cite{sengun2019analyzing} \\
    Tribunal dataset & \xmark & report & 2M & chat msgs. & LoL & \cite{neto2017studying} \\
    Tribunal dataset & \xmark & report & 11M & chat msgs. & LoL & \cite{blackburn2014stfu} \\
    CONDA & \xmark & message & 50k & chat msgs. & DOTA & \cite{weld2021conda} \\
    For honor dataset & \xmark & chat & 1800 & chat msgs. & For Honor & \cite{canossa2021honor} \\

    \bottomrule
  \end{tabular}%
  \vspace{-5mm}
\end{table}

\subsection{Lessons Learned and Implications}
\label{sssec:implications}
\noindent
First, our SLR showed that, despite a great interest in this topic, few papers propose and implement (and share) solutions to address the problem of toxicity and harassment in video games. Such a finding should serve as a call for action.

Second, our open-source \datasetname{} serves as a foundation for future research in this domain. Thanks to \datasetname{}, it is possible to develop, or test, novel countermeasures. Indeed, when we begun this study, we were surprised to find a lack of publicly-available resources that are {\small \textit{(i)}}~game-specific and {\small \textit{(ii)}}~labeled at the message level. Such a lack supports our hypothesis that the panorama of practical countermeasures to toxicity/harassment is limited from a research viewpoint.

Third, and anecdotally, while testing IGC-BERT on YouTube (in §\ref{ssec:youtube}) we attempted to test it also on videos that, in our opinion, did not have any sort of toxic behavior. So, we downloaded the captions taken from some videos meant for kids (Specifically, we used: \href{https://www.youtube.com/watch?v=zD2SxyMYX0U}{Caillou at the Restaurant}) We were surprised when we noticed that our model found instances of toxicity. Upon checking, we found that the following lines had been flagged as toxic: ``what's a donkey doing here'' and ``your toy when your dad gets back''. We find such a result fascinating: such lines, within a kids' show, are clearly free of any toxic content; however, by hypothesizing that such text is sent during a LoL match, the meaning substantially changes. Our takeaway is that such findings demonstrate the challenges, but also the need, of developing context-specific detectors of toxicity. This result further confirms our claim: ultimately, \textbf{our contributions are designed for LoL. It is unrealistic to expect effectiveness in other games.}\footnote{In our repository~\cite{repository}, we have evaluated our considered models also on a dataset of 3k Dota2 messages (as also done in~\cite{du2024automatic}) and on the 1k messages in YouToxic (as also done in~\cite{miok2019prediction}). These experiments are orthogonal to our main objective. However, our findings confirm our hypothesis: our IGC-BERT does not work so well in contexts different from LoL. We consider this empirical finding crucial for future work.} Such an objective can, however, be pursued via ensembles (e.g., developing multiple game-specific detectors, and using the one specific for the given context).

Lastly, integrating our IGC-BERT model in a browser extension working locally presents tradeoffs. These can be mitigated by using, e.g., model distillation~\cite{hsieh2023distilling} to produce a ``smaller'' model that can work faster, but at the expense of detection performance. Such engineering endeavours are beyond our scope.

\subsection{Ethical Considerations}
\label{sssec:ethics}
\noindent
When we carried out our research, our institutions did not have any formal IRB process. However, we performed our study according to best practices~\cite{bailey2012menlo}. 

For our \datasetname{} dataset, we relied on the expertise of 8 veteran LoL players. Their participation in such an effort was voluntary, and they willingly agreed to contribute to this study. Participants were aware that their actions would be used to produce a dataset used for research purposes, which was planned to be publicly shared afterwards. We did not collect any sensitive or personally-identifiable information about our participants~\cite{sensitiveEU,sensitiveUSA}. To preserve the anonymity of our participants, we cannot disclose additional details about them.

We do not envision any potential negative outcome from our study. We warned that our resources are prototypes, and should not be used to, e.g., claim that any given user is exhibiting toxic behavior without any additional form of validation. In contrast, openly releasing our tools outweighs any risk, since it is the best way to address the widespread problem of toxicity and harassment in video games: sharing our resources enables future work to build upon our findings, fostering more (and much needed) research in this domain.
\section{Conclusions and Future Work}
\label{sec:conclusions}
\noindent
Despite the benefits that video-games bring to our lives, thousands of players still report being victim of targeted toxicity/harassment. This is particularly true in competitive multiplayer titles, where the goal is not always limited to ``playing for fun'', leading to players forgetting that there are human beings on the other side of the screen, who can be severely hurt by reading certain messages.

We tackled this problem and found that, unfortunately, there are no datasets with labeled instances of toxic messages exchanged during a match. So, we created a novel dataset, \datasetname{}, thanks to the contributions of 8 veteran players of the ever-popular LoL game. We used \datasetname{} to carry out a number of experiments, such as developing an ML-based model that outperforms general-purpose toxicity detectors, testing the model on YouTube videos, and integrating the model on a browser extension. Altogether, our resources should inspire future work focused on practical solutions to the problem of toxicity and harassment in video games. We reached out to RIOT to inform them of our solutions.

We identify three avenues for future work. First, expanding our \datasetname{} dataset by labeling additional instances taken by the Tribunal dataset, or testing our IGC-BERT model on messages in the Tribunal dataset not included in \datasetname{}. Second, the development of datasets (or models) focused on different games than LoL. Third, the integration of our resources with orthogonal detection techniques, such as those based on gameplay/behavioral elements (e.g., correlating chatlogs with ``repeated pings'' or other types of griefing behavior~\cite{paul2015enjoyment}), that can further augment the identification of toxic behavior during a match.

\textbf{Acknowledgments.} We would like to thank the anonymous ESORICS'26 reviewers for the great feedback. Parts of this research has been funded by Hilti.

{\scriptsize
\bibliographystyle{splncs04}

}

\appendix


\begin{figure}[b]
    \centering
    \includegraphics[width=0.5\linewidth]{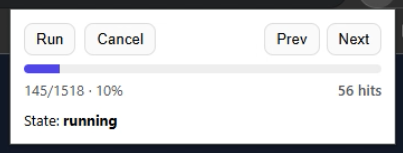}
    \vspace{-3mm}
    \caption{The graphical interface of our browser extension. The analysis can be stopped or resumed.}
    \label{fig:interface}
    \vspace{-7mm}
\end{figure}

\end{document}